\documentclass[aps,prd,superscriptaddress,amsfonts,amssymb,amsmath,eqsecnum,nofootinbib,twocolumn,floatfix]{revtex4}
 \usepackage{color,graphicx}
 \usepackage[utf8]{inputenc}
 \usepackage[T1]{fontenc}
 \usepackage[english]{babel}
 \definecolor{darkblue}{rgb}{0,0,0.7}
\definecolor{darkred}{rgb}{0.7,0,0}
\definecolor{darkgreen}{rgb}{0,0.4,0}
\usepackage[unicode, colorlinks, citecolor=darkblue, linkcolor=darkred, urlcolor=blue]{hyperref}

\allowdisplaybreaks
\begin{document}

\author{Alexandr Karpenko}
\affiliation{Faculty of Physics, M.V. Lomonosov Moscow State University, Leninskie Gory, Moscow 119991,  Russia}

\author{Sergey P. Vyatchanin}
\affiliation{Faculty of Physics, M.V. Lomonosov Moscow State University, Leninskie Gory, Moscow 119991,  Russia\\
     Quantum Technology Centre, M.V. Lomonosov Moscow State University, Leninskie Gory, Moscow 119991,  Russia}

\date{\today}
	
\title{Dissipative coupling, dispersive coupling and their combination \\  in the cavity-less opto-mechanical systems}

\begin{abstract}
The approach of variational measurement is applied to the cavity-less dissipative coupling (when test mass displacement is changing transmittance of a single mirror). The result obtained is compared to the cavity-less dispersive coupling (when the mass of displaced single mirror is changing the phase in reflected wave). These types of coupling are compared for ponderomotive squeezing. In addition, the cavity-less variant of combined dissipative and dispersive coupling is explored. It is established that combined coupling results in a stable optical rigidity even in case of a single pump. Finally, it is confirmed that variational measurement can be applied to combined coupling.

\end{abstract}

\maketitle

\section{Introduction}
Opto-mechanics is studying interactions developing between the light inside an optical cavity and a mechanical oscillator or a free mass \cite{AspelmeyerRMP2014}. A simple realization of the well known dispersive opto-mechanic coupling is presented by a cavity where position of a mechanical body (moving mirror) is changing the cavity eigen frequency with the light pressure generating a force proportional to the optical power or to the number of optical quanta inside the optical cavity.  Opto-mechanical systems having several degrees of freedom feature more complex interactions ranging from radiation puling (negative radiation pressure) \cite{Povinelli05ol,maslov13pra}, opto-mechanical interaction proportional to the quadrature of electromagnetic field \cite{93a1VyMaJETP,96a1VyMaJETP,matsko97apb,02a1KiLeMaThVyPRD} to the interaction depending on velocity (rather than coordinate) of the test mass in a mechanical system. \cite{90BrKhPLA,00a1BrGoKhThPRD}.

Accounting for opto-mechanics effects is especially important in precision measurements 
in gravitational wave detectors \cite{aLIGO2013,aLIGO2015,MartynovPRD16,AserneseCQG15, DooleyCQG16,AsoPRD13}, in torque sensors \cite{WuPRX2014, KimNC2016, AhnNT2020}, and in magnetometers \cite{ForstnerPRL2012, LiOptica2018}. Optomechanical systems can be also used for frequency conversion between microwaves and light \cite{LambertAQT2019, LaukQST2020} or for testing the limits of quantum mechanics \cite{BassiRMP2013, ArndtNP2014}. 

The accuracy in measuring mechanical position in an opto-mechanical system is usually limited due to the quantum back action known as standard quantum limit (SQL)  \cite{Braginsky68,BrKh92}. SQL was studied in many systems ranging from macroscopic kilometer-size gravitational wave detectors \cite{02a1KiLeMaThVyPRD} to microcavities \cite{Kippenberg08,DobrindtPRL2010}. An example of measurements restricted by SQL is provided by detecting classical force acting on a test mass in opto-mechanical systems. At the same time SQL does not pose a fundamentally unavoidable limit for the measuring force. It can be avoided by applying a variational measurement \cite{93a1VyMaJETP, 95a1VyZuPLA, 02a1KiLeMaThVyPRD}, or squeezed light input \cite{LigoNatPh11, LigoNatPhot13, TsePRL19, AsernesePRL19, YapNatPhot20, YuArxiv20, CripeNat19}, or opto-mechanical velocity measurement \cite{90BrKhPLA,00a1BrGoKhThPRD}, or optical rigidity \cite{99a1BrKhPLA, 01a1KhPLA}. 
SQL can be surpassed also with coherent quantum noise cancellation \cite{TsangPRL2010, PolzikAdPh2014, MollerNature2017}.

Opto-mechanic coupling can be broken into two types, namely: into dispersive and dissipative coupling. In dispersive coupling displacing mirror is changing normal cavity frequency, whereas in dissipative coupling displacing test mass brings about a change in the input mirror transparency altering, thereby, the cavity relaxation rate. Dissipative coupling was initially proposed theoretically \cite{ElstePRL2009} and confirmed experimentally \cite{LiPRL2009,WeissNJP2013,WuPRX2014, HryciwOpt2015} nearly a decade ago. The phenomenon was studied in numerous opto-mechanical systems, including the Fabry-Perot interferometer \cite{LiPRL2009,WeissNJP2013,WuPRX2014, HryciwOpt2015}, the Michelson-Sagnac interferometer \cite{XuerebPRL2011, TarabrinPRA2013, SawadskyPRL2015}, and ring resonators \cite{HuangPRA2010,HuangPRA2010b}. It was shown  that an opto-mechanical transducer based on dissipative coupling between optical and mechanical degrees of freedom allows realizing quantum speed meter which, in turn, helps avoid SQL \cite{16a1PRAVyMa}.

The natural question is whether combination of dispersive and dissipative couplings could improve sensitivity of optomechanical system to signal displacement of test mass or vice versa? In general case the problem is rather complex.  In this paper we analyze simplified model of dispersive, dissipative coupling and their combination in the cavity-less opto-mechanical system. We show that squeezed quadratures  in case of pure dispersive and pure dissipative coupling are strongly different, so it looks that combination (``mariage'') of dispersive and dissipative coupling do not promise any profit. However, we show that applying variational measurement to a combined coupling is possible and allows to surpass SQL.

Dispersive coupling inside the cavity makes the normal frequency dependent on position of the test mass. For example, in the Fabry-Perot cavity it is the position of the input or the end mirror. Therefore, we can model dispersive coupling a cavity-less optomechanical system by a stand-alone moving mirror (acting as a free test mass) having a phase of reflected light dependent on the test mass position. 

\begin{figure}
\includegraphics[width=0.47\textwidth]{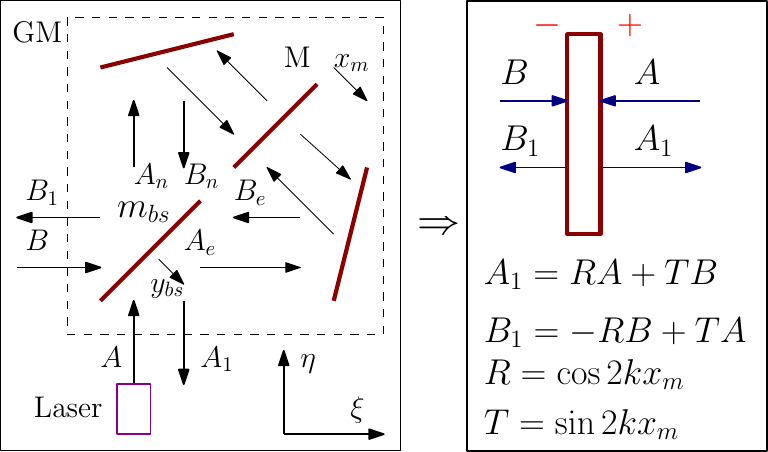}
 \caption{When BS position is fixed ($y=0$) the Michelson-Sagnac interferometer becomes a GM having transparency and reflectivity depending on position of the test mass (a completely reflecting mirror $M$) giving an example of dissipative coupling. In the opposite case of moving BS and fixed position  $x_m=0$ of mirror $M$ the model of a mirror having both dispersive and dissipative coupling is obtained.}\label{MSI}
\end{figure}

In turn, dissipative coupling inside the cavity means that its relaxation rate depends on position of the test mass which --- in case of the Fabry Perot cavity --- implies that transmittance of the input mirror depends on position of the test mass. Thus, we can model dissipative coupling in a cavity-less optomechanical system as a mirror having amplitude reflectivity $R$ and transmittance $T$ varying with position of the test mass. Specifically, this corresponds to the Michelson-Sagnac interferometer (MSI)  \cite{XuerebPRL2011, TarabrinPRA2013, SawadskyPRL2015} acting as a generalized mirror (GM) where the test mass is represented by a moving completely reflecting mirror $M$ (see Fig.~\ref{MSI}).

Ovbiously, in cavity-less optomechanical system dispersive coupling can be also called as a phase coupling and dissipative one — as amplitude coupling. However, below we call them as dispersive and dissipative couplings to underline their connection with cavity optomechanical system.

We also examine the opposite case of a moving beam splitter (BS) in MSI and mirror $M$ ($x_m$ being constant) having a fixed position described by a model of mirror having combined (dispersive and dissipative) coupling. 

\section{MSI as a generalized mirror}

MSI consists of a 50/50 beam splitter and three mirrors (we consider them opaque). This interferometer can be considered as a GM having amplitude transmittance $T$ and reflectivity $R$ depending on displacement $x_m,\ y_{bs}$. This dependence is originated by interference of reflected from mirror $M$ waves $B_n,\ B_e$ (see Fig.~\ref{MSI}) on BS. So input-output relations has form:
\begin{subequations}
  \label{MSIA1}
  \begin{align}
    \label{MSIA1a}
   \mathcal A_1 & =  \mathcal B T  + e^{-ik\sqrt 2y_{bs}} \mathcal A R, \\
    \label{MSIA1b}
   \mathcal B_1 &=  \mathcal A  T - e^{ik\sqrt 2y_{bs}}\mathcal B R,\\
      R &=   \cos k\big(2x_m +\sqrt 2 y_{bs}\big), \ T=  \sin k\big(2x_m +\sqrt 2 y_{bs}\big).\nonumber
   \end{align}
   \end{subequations}
The fact that MSI behaves like a GM can be seen from equations \eqref{MSIA1} and \eqref{M1}, which relate the input and output fields for MSI and a normal mirror.

A detailed analysis of MSI is given in Appendix~\ref{appMSI}. When analyzing MSI, we assume that the waves passing through BS acquire a phase shift equal to $\frac{\pi}{2}$, and the phase of the waves reflected from BS is determined only by the displacement of the BS itself (see Eqs. \ref{initMSI}). We also assume that the spectral frequencies $\Omega$, characterizing the displacements of BS and the moving mirror $M$ are low enough: $\Omega t_{in}\ll 1$, where $t_{in}$ is round trip time of light between BS and mirror $M$. This means that the circulating fields change phase almost instantly at small displacements of BS and $M$.

 From \eqref{MSIA1} one can see that in case of fixed BS ($y_{bs}$ is a constant) and movable mirror $M$ we have amplitude transmittance $T$ and reflectivity $R$ depending on displacement $x_m$, and exponents  in (\ref{MSIA1a}, \ref{MSIA1b} are constants --- it means pure dissipative coupling. 

In alternative case of ``synchronized'' displacements of BS and mirror $M$, so that $x_m + \sqrt 2 y_{bs}$ is a constant, reflectivity $R$ and transmittance $T$ are constants, but exponents in (\ref{MSIA1a}, \ref{MSIA1b}, describing phase shifts of reflected waves, depend on BS displacement --- it corresponds to pure dispersive coupling.

In general case MSI can demonstrate both dispersive and dissipative coupling.

 We can designate displacement as 
   \begin{align}
   \label{xy}
   x_m &= x_0 + x ,\ y_{bs}= y_0 + y
  \end{align}
 where $x_0,\ y_0$ are the mean constants (to be chosen) and $x ,\ y$ are small variables.  Then we can expand $R,\ T$ \eqref{MSIA1} into a series
  \begin{subequations}
  \label{expRT}
  \begin{align}
   \label{Rexp}
    R &  \simeq R_0 + T_0\, k\big[2x + \sqrt 2\, y\big] \, , \\
     \label{Texp}
    T &  \simeq -T_0 + R_0\, k\big[2x + \sqrt 2\, y\big]\,,\\
    R_0 &= \cos k  \big(2x_0 +\sqrt 2\, y_0\big),\\
    T_0 &= - \sin k  \big(2x_0 +\sqrt 2\, y_0\big),\nonumber
 \end{align}
 \end{subequations}
 where $k=\omega_0/c$, $\omega_0$ is the carrier frequency of light waves.
 For simplicity we put $y_0=0$ below, then only $x_0$ defines $T_0,\ R_0$.
 
 We express amplitudes of waves as a large constant amplitudes (denoted by capital letters) plus small amplitudes (denoted by the same letters in low-case) to describe noise and signal components. For example
 \begin{align}
  \label{expA}
  \mathcal A & = A + \hat a,\quad \mathcal A_1  = A_1 + \hat a_1,\quad \text{and so on.}
 \end{align}
We choose the normalization making the mean pump power of incident wave equal to $\hslash \omega_0 |A|^2$ \cite{02a1KiLeMaThVyPRD}.  
Here and below we assume that input wave is in coherent state 
so operators $\hat a,\ \hat b$ describe the vacuum fluctuation wave having the following commutator and correlator
\begin{align}
  \label{comm}
  \left[\hat a(t), \hat a^\dag(t')\right] &=  \left[\hat b(t), \hat b^\dag(t')\right]= \delta(t-t'),\\
  \label{corr}
  \left\langle\hat a(t) \hat a^\dag(t')\right\rangle &=\left\langle\hat b(t) \hat b^\dag(t')\right\rangle= \delta(t-t')
\end{align}

The Fourier transform can be defined as follows
\begin{align}
 \hat a(t) &= \int_{-\infty}^\infty a(\Omega) \, e^{-i\Omega t}\, \frac{d\Omega}{2\pi}
\end{align}
and similarly for other values denoting the Fourier transform by the same letter without hat. One can derive the following from \eqref{comm} and\eqref{corr} for the Fourier transform of the input fluctuation operators:
 \begin{align}
  \label{comm1}
  \left[ a(\Omega),  a^\dag(\Omega')\right] &=\left[ b(\Omega),  b^\dag(\Omega')\right]= 2\pi\,\delta(\Omega -\Omega'),\\
  \label{corr1}
  \left\langle a(\Omega)  a^\dag(\Omega')\right\rangle &= \left\langle b(\Omega)  b^\dag(\Omega')\right\rangle =2\pi\, \delta(\Omega -\Omega')
\end{align}

\section{The cavity-less dissipative coupling}

Let us consider a particular case when BS position is fixed ($y_0=0,\ y=0$), then MSI is  a GM modeling dissipative coupling \cite{XuerebPRL2011}: reflectivity and transmittance depend on position of mirror $M$ (test mass). 

Let us consider the simplest case of a single pump where the mean amplitudes are
\begin{align}
 \label{oneP}
 B&=0,\quad A=A^*\,.
\end{align}
Then using \eqref{MSIA1}, \eqref{expRT} and \eqref{expA} we can write down for the small amplitudes:
\begin{subequations}
 \label{a1b1}
 \begin{align}
 \hat a_1 &=-T_0 \hat b + R_0 \hat a +AT_0\, 2kx,\\
 \hat b_1 &= -T_0\hat a - R_0 \hat b + A R_0\, 2kx
\end{align}
\end{subequations}

Let us introduce the quadratures in the frequency domain:
 \begin{align}
 \label{a1b1quad}
  a_a &= \frac{a+ a_-^\dag}{\sqrt 2},\quad a_p = \frac{a - a_-^\dag}{i\sqrt 2},\quad a_-\equiv a(-\Omega)
 \end{align}
 For other small amplitudes the quadratures are defined in a similar manner. 
 
 We can rewrite \eqref{a1b1} for quadratures in the frequency domain
 \begin{subequations}
  \label{a1b1quad2}
  \begin{align}
  \label{a1a}
  a_{1a} &=-T_0  b_a + R_0  a_{a} +\sqrt 2AT_0\, 2kx(\Omega),\\
  \label{a1p}
  a_{1p} &= -T_0  b_p + R_0  a_p ,\\
  \label{b1a}
  b_{1a} &=  -T_0 a_a - R_0  b_a + \sqrt 2A R_0\, 2kx(\Omega),\\
  \label{b1p}
  b_{1p} &= - T_0 a_p - R_0  b_p 
\end{align}
\end{subequations}
These equations indicate on presence of dissipative coupling as the information on displacement is given by the {\em amplitude}  quadratures in reflected and transmitted waves. In contrast, for dispersive coupling information on displacement is provided only by the phase quadrature in reflected wave. It is shown in Sec.~\ref{dispC} below.

Signal $F_s$ and the fluctuation back action force \eqref{FbaM} interact with free test mass $m$ (the mass of mirror $M$). For a particular case \eqref{oneP} we obtain the following in the frequency domain:
\begin{align}
  \label{xdiss}
 - m \Omega^2 x = 2\sqrt 2 \, \hslash k A\, b_p+ F_s 
\end{align}
If $T_0\ll R_0$  we obtain combining (\ref{b1a}, \ref{b1p}, \ref{xdiss})\footnote{
Strictly speaking we have to take the linear combination:
\begin{subequations}
 \begin{align}
  \tilde c_{1a} &= R_0 b_{1a} +T_0 a_{1a},\\ 
  \tilde c_{1p} &= - T_0 b_{1p} + R_0 a_{1p}.
 \end{align}
\end{subequations}
In case $T_0\to 0$ it turns into \eqref{b1ap}. The accurate consideration is given in Sec.~\ref{combC}.
}
\begin{subequations}
 \label{b1ap}
\begin{align}
 b_{1a} & \simeq -b_a - \mathcal K \cdot b_p -\sqrt{2\mathcal K}\cdot f_s, \\
 \label{KK}
 b_{1p} &\simeq -b_p, \quad \mathcal K= \frac{8\hslash k^2 A^2}{m\Omega^2},\quad 
               f_s=\frac{F_s(\Omega)}{\sqrt{2\hslash m\Omega^2}}
\end{align}
\end{subequations}
Here $\mathcal K$ is the recalculated pump power, $f_s$ is the signal force normalized to SQL. 
We see that information on back action and the signal force contains in the amplitude quadrature and in the phase quadrature remains the same.

Equation \eqref{b1ap} describes ponderomotive squeezing which allows to surpass SQL. To realize it one can apply the variation measurement approach \cite{93a1VyMaJETP, 95a1VyZuPLA, 02a1KiLeMaThVyPRD} to compensate for the back action. This can be accomplished through measuring a combination of amplitude and phase quadratures in transmitted wave using homodyne detection:
\begin{subequations}
 \label{btheta}
  \begin{align}
  b_\theta &= b_{1a} \cos\theta + b_{1p}\sin\theta =\\
		=& - b_a \cos\theta - b_p\left(\sin\theta +\mathcal K\cos\theta\right)  -\cos\theta\sqrt{2\mathcal K}\cdot f_s,\nonumber
  \end{align}
\end{subequations}
where $\theta$ is the homodyne angle. Choosing 
\begin{align}
 \label{tan}
 \tan\theta = - \mathcal K (\Omega_0)
\end{align}
 we compensate for the back action completely but only at a chosen frequency $\Omega_0$.

Let input fields are in coherent state. This means that correlators \eqref{corr1} are valid and single-sided power spectral densities (PSD) of quadratures are equal to $S_a(\Omega) =S_p(\Omega)=1$ \cite{02a1KiLeMaThVyPRD}. Then PSD of noise recalculated to $f_s$ can be easily derived from \eqref{btheta}:
\begin{align}
 \label{Sfs}
 S_{fs}^\text{diss} = \frac{1}{2\mathcal K} + \frac{(\mathcal K_0 - \mathcal K)^2}{2\mathcal K},\quad
	 \mathcal K_0 =\mathcal K(\Omega_0)
\end{align}
Here $S_{fs}^\text{diss}=1$ corresponds to SQL sensitivity\footnote{
Strictly speaking the PSD minimum \eqref{Sfs} is achieved not at condition \eqref{tan} but when  
$ \tan\theta = - \sqrt{\mathcal K^2 (\Omega_0) -1}$.
However, in limit $\mathcal K(\Omega_0)\gg 1$ both conditions coincide and below we make use of condition \eqref{tan}.
}. 
In case $\mathcal K_0\gg 1$ the minimum PSD $S_{fs}^\text{min}\simeq \frac{1}{2\mathcal K_0}$ is realized inside a narrow bandwidth $\Gamma$:
\begin{align}
 \label{CRb}
 \frac{\Gamma}{\Omega_0} \simeq 2S_{fs}^\text{min}
\end{align}
Here $\Gamma$ is defined as $S_{fs}^\text{diss}(\Omega_0 \pm \Gamma/2) \simeq 2 S_{fs}^\text{min}$. The relation \eqref{CRb} corresponds to the known Cramer-Rao bound \cite{Mizuno,Mizuno93, Miao2017}.

\begin{figure}
\includegraphics[width=0.2\textwidth]{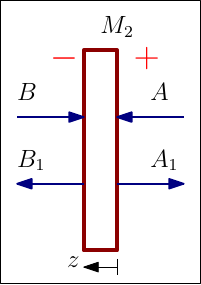}
 \caption{The cavity-less dispersive coupling. A moving mirror having amplitude reflectivity $R_0$ and transmittance 		$T_0$ is a free test mass; its position is measured through detecting phase (the phase quadrature) in reflected wave. }\label{Disp}
\end{figure}
  
\section{The cavity-less dispersive coupling}\label{dispC}

Let us  consider the simplest case of dispersive coupling. Moving mirror $M_2$ having amplitude reflectivity $R_2$ and transmittance 		$T_2$ is a free test mass; its position $z$ is measured by detecting phase (the phase quadrature) in reflected wave, see Fig.~\ref{Disp}. Again we consider the case \eqref{oneP} (i.e., $B=0$ and  $A$ is real). This scheme is widely known (e.g., \cite{96a1VyMaZu}) and one can write down the following for the output small amplitudes
\begin{subequations}
 \label{a1b1Disp}
 \begin{align}
 \hat a_1 &=T_2 \hat b + R_2 \hat a + AR_2\, 2i kz,\\
 \hat b_1 &= T_2\hat a - R_2 \hat b \,.
\end{align}
\end{subequations}
We rewrite \eqref{a1b1Disp} for quadratures in the frequency domain 
\begin{subequations}
 \label{a1b1Disp2}
 \begin{align}
  a_{1a} &=T_2  b_a + R_2  a_{a} ,\\
  \label{a1b1Disp2b}
  a_{1p} &=T_2  b_p + R_2  a_p + \sqrt 2 AR_2\, 2kz(\Omega),\\
  b_{1a} &= T_2 a_a - R_2  b_a ,\\
  b_{1p} &= T_2 a_p - R_2  b_p 
\end{align}
\end{subequations}
 We can notice that information on displacement is contained only in phase quadrature $a_{1p}$ of reflected wave.
  
Signal $F_s$ and fluctuation back action force \eqref{Fm1} affect the free test mass $m_2$ (it is the mass of moving mirror $M_2$). In case \eqref{oneP} we obtain for the frequency domain:
 \begin{align}
  \label{m2disp}
  -m_2\Omega^2 z = 2\sqrt 2 \hslash k R_2 A\left(R_2 a_{a} + T_2 b_a\frac{}{}\right) + F_s
 \end{align}
We don't take into account in \eqref{m2disp} the term proportional to $A^2$, because it is responsible for a constant force, which can be compensated by applying the same modulus of force to the mirror $M_2$ in the opposite direction.

Substituting \eqref{m2disp} into \eqref{a1b1Disp2} we obtain
\begin{subequations}
 \label{a1ap}
 \begin{align}
  a_{1a} &=T_2  b_a + R_2  a_{a} ,\\
  a_{1p} &=T_2  b_p + R_2  a_p - \mathcal N \, a_{1a} -\sqrt{2\mathcal N}\,f_s,\\
         & \mathcal N =\frac{8\hslash k^2 R_2^2 A^2}{m_2\Omega^2}
 \end{align}
\end{subequations}
Obviously, the output quadratures $b_{1a},\ b_{1p}$ of transmitted wave do not contain any information on displacement $z$. 

We see that equations \eqref{b1ap} for dissipative coupling are similar to equations \eqref{a1ap} obtained for dispersive coupling. The only difference deals with the phase quadrature  $a_{a1}$ containing information on displacement in dispersive coupling \eqref{a1ap} being replaced by the amplitude quadrature $b_{1a}$ for dissipative coupling \eqref{b1ap}. 

The formula \eqref{Sfs} is also valid for dispersive coupling having different homodyne angle: $\tan\theta= 1/\mathcal N(\Omega_0)$.

\begin{figure}
\includegraphics[width=0.47\textwidth]{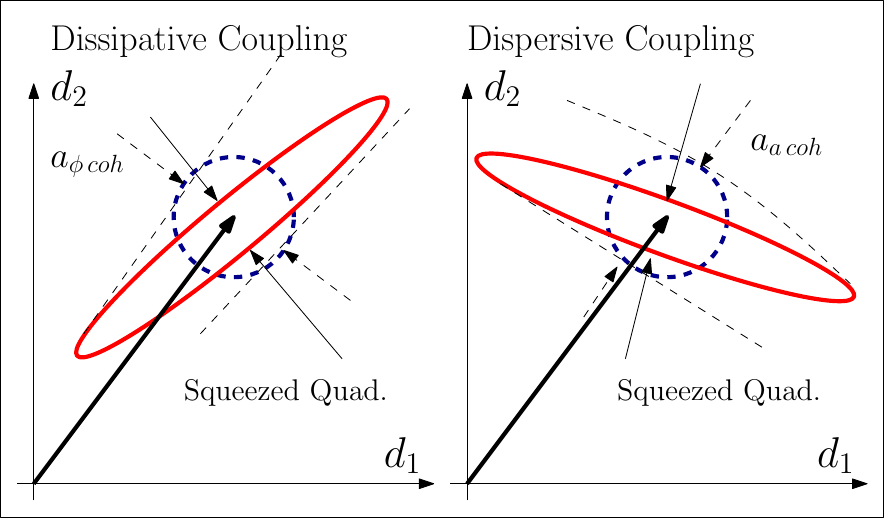}
 \caption{Squeezing in reflected wave on the phase plane is denoted by ellipses plotted for dissipative and dispersive coupling. Incident wave is in coherent state, its fluctuations are expressed using dotted circles. Axis $d_1,\ d_2$ define coordinates on phase plane.
 }\label{Pplane}
\end{figure}

It is obvious that both transforms \eqref{b1ap} and \eqref{a1ap} describe squeezing. The difference is illustrated in Fig.~\ref{Pplane}. Here incident waves are assumed to be in coherent state which is presented as mean amplitude (havier arrow) plus fluctuations depicted by dotted circles with equal variation of amplitude $a_{a\, coh}$ and phase $a_{\phi\, coh}$ quadratures. For dissipative coupling \eqref{b1ap} $a_{a\, coh}= b_a,\ a_{\phi\, coh}= b_p$, for dispersive one \eqref{a1ap} $a_{a\, coh}= T_2  b_a + R_2  a_{a},\ a_{\phi\, coh}= T_2  b_p + R_2  a_{p}$.
The reflected wave is squeezed (fluctuations have an elliptical shape): for dispersive coupling amplitude quadrature does not change and the phase quadrature is  increasing whereas for dissipative coupling the phase quadrature remains the same and the amplitude quadrature increases.

\section{Combined coupling}\label{combC}

Let us consider {\em combined} coupling when both dissipative and dispersive coupling are present.
We will proceed with analyzing the same MSI as shown in Fig.~\ref{MSI} but with moving BS; 
it has a test mass $m_{bs}$ and the coordinate is $y$ \eqref{xy} for a fixed mirror $M$ ($x=0$). 
Then using \eqref{MSIA1}, \eqref{expRT} and \eqref{expA} we obtain for small amplitudes:
\begin{subequations}
  \label{MSIA1c}
  \begin{align}
   a_1 & = - T_0 b +R_0 a + \\
		  &\quad +R_0 B \sqrt 2k y +T_0 A \sqrt 2k y  - R_0 A\,ik\sqrt 2y , \\
  b_1 &=  - T_0 a -R_0 b +\\
		  &\quad + R_0A\,k\sqrt 2y -T_0B\,k\sqrt 2y - R_0 B ik\sqrt 2y\,.
  \end{align}
 \end{subequations}
 In case of one pump \eqref{oneP} we rewrite input-output relations for quadratures in the frequency domain as follows:
\begin{subequations}
  \label{MSIA1cquad}
  \begin{align}
  \label{MSIA1ca1a}
   a_{1a} & = - T_0 b_a +R_0 a_a  + T_0 A  2k y   , \\
   \label{MSIA1a1p}
   a_{1p} & =  - T_0 b_p +R_0 a_p  - R_0 A\,2k y,\\
   \label{MSIA1cb1a}
   b_{1a} &=  - T_0 a_a -R_0 b_a 	+ R_0A\,2k y  \,, \\
   \label{MSIA1cb1p}
   b_{1p} &=  - T_0 a_p -R_0 b_p  \,.
  \end{align}
 \end{subequations}
 We see that both dissipative and dispersive coupling are present.
 Indeed, coordinate term in \eqref{MSIA1a1p} corresponds to dispersive coupling (compare with \eqref{a1b1Disp2b}), whereas coordinate terms in (\ref{MSIA1ca1a}, \ref{MSIA1cb1a}) are describing dissipative coupling (compare with (\ref{a1a}, \ref{b1a})). 
 
 Equations \eqref{MSIA1cquad} should be supplemented by an equation describing mechanical degree of freedom. In approximation \eqref{expRT} and \eqref{oneP} we can express coordinate $y$ in the frequency domain using \eqref{FyBS}, taking into account only the first order of smallness by the y offset. The  result is as follows:
  \begin{align}
   \label{y}
   y &=   \frac{2\hslash k A \left(-R_0^2 a_a + R_0T_0 b_a + b_p\right)}{(K - m_{bs}\Omega^2)}+\\
   \label{K}
	 &\quad +\frac{\sqrt{2\hbar m_{bs}\Omega^2}}{(K - m_{bs}\Omega^2)}\cdot f_s,\\
	 \label{Kfs}
    &  K= 4\hslash k^2 A^2 R_0T_0, \quad f_s=\frac{F_s}{\sqrt{2\hslash m_{bs}\Omega^2}},
  \end{align}
where $f_s$ is the signal force normalized to SQL, $K$ is the optical rigidity which appears due to presence of both dissipative and dispersive coupling\footnote{Indeed, for single pump ($\mathcal B=0$) the Lebedev light pressure force is equal to \eqref{FyBS1} ($F_{bs} \sim -R^2 |\mathcal A|^2$), this term is originated by dispersive coupling. The rigidity appears due to dependence $R(y)$ (dissipative coupling).}. The rigidity is {\em positive} when $T_0R_0 >0$, see definition \eqref{expRT}.  Note, the rigidity $K$ is a constant,  it does not depend on frequency\footnote{
Strictly speaking, the accurate account of the Doppler effect provides for a weak viscosity \cite{96a1VyMaZu} (about $\sim F_{lp}/c$, $F_{lp}$ is a constant light pressure force, $c$ is the light velocity). Yet, here we can neglect it.}.    

Terms $\sim a_a,\ b_a$ in \eqref{y} correspond to back action force of dispersive coupling, whereas term $\sim b_p$ in \eqref{y} is relevant to back action force of dissipative coupling.

In order to apply variation measurement we have to generalize this case over two output beams. 
It is possible to measure arbitrary quadratures in transmitted and reflected waves using homodyne detection and taking weighted sum as a result. It means that we can take an arbitrary linear combination of quadratures \eqref{MSIA1cquad}. We can optimize coefficients in this combination in order to find the PSD minimum recalculated to $f_s$ \eqref{b1ap} at a certain predefined frequency $\Omega_0$ :
 \begin{subequations}
   \label{SfsMin}
 \begin{align}
 S_{fs}^\text{comb}(\Omega) & = \frac{1}{2LR_0}\times\\
	 &\times \left\{ \frac{R_0^2+1}{T_0}\left[ \frac{L_0 -L}{L_0-1}\right ]^2 
		  + \frac{T_0(L-1)^2}{\big(R_0^2+1\big)}  \right\},\nonumber\\
        \label{L0}
		L &= \frac{K}{m_{bs}\Omega^2},\quad L_0= \frac{K}{m_{bs}\Omega_0^2}
 \end{align}
 \end{subequations}
See details in Appendix \ref{derivation}. Recall that $S_{fs}^\text{comb}=1$ corresponds to SQL. When $\Omega=\Omega_0$ the first term in \eqref{SfsMin} is equal to zero and the second term defines the minimum $S_{fs}^\text{comb}(\Omega_0)$. However, when $\Omega = \Omega_0 + \Delta\Omega$ the first term is dramatically increasing defining the effective bandwidth $\Gamma$. 
 Requiring a twofold increase in PSD at $\Delta\Omega=\Gamma/2$ we find: 
 \begin{align}
  \label{DeltaOmega}
  \frac{\Gamma}{\Omega_0} &\simeq \frac{T_0 (L_0-1)^2}{(R_0^2+1)L_0}\,,\quad 
     \frac{\Gamma}{\Omega_0} \simeq 2 R_0 S_{fs}^{min} (\Omega_0)
 \end{align}
 Here relation between the bandwidth $\Gamma$ and $S_{fs}^{min}$ is similar to \eqref{CRb} corresponding to the Cramer-Rao bound \cite{Mizuno,Mizuno93, Miao2017}.
 
\begin{figure*}
\includegraphics[width=0.7\textwidth]{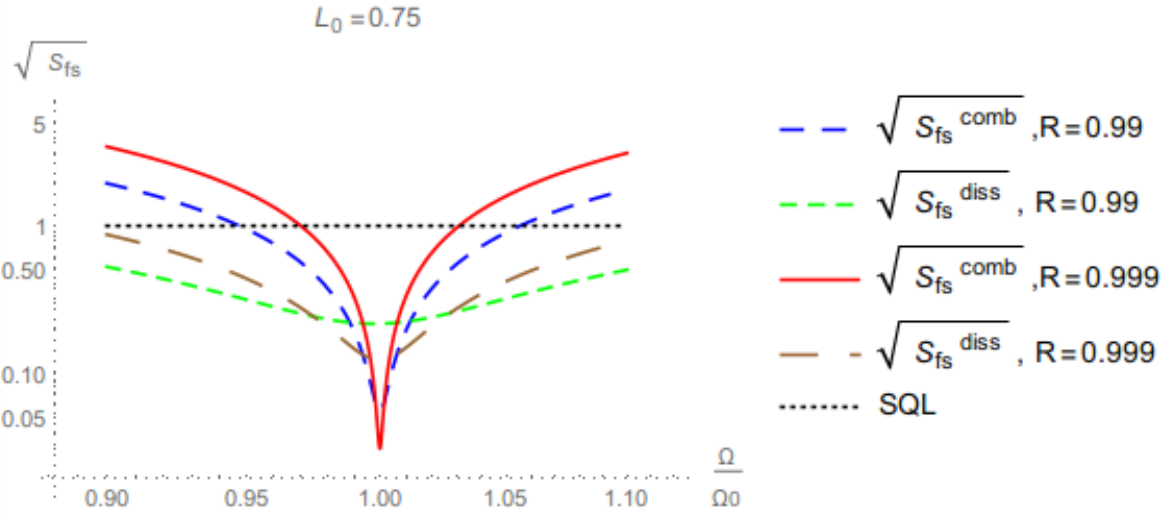}
\includegraphics[width=0.7\textwidth]{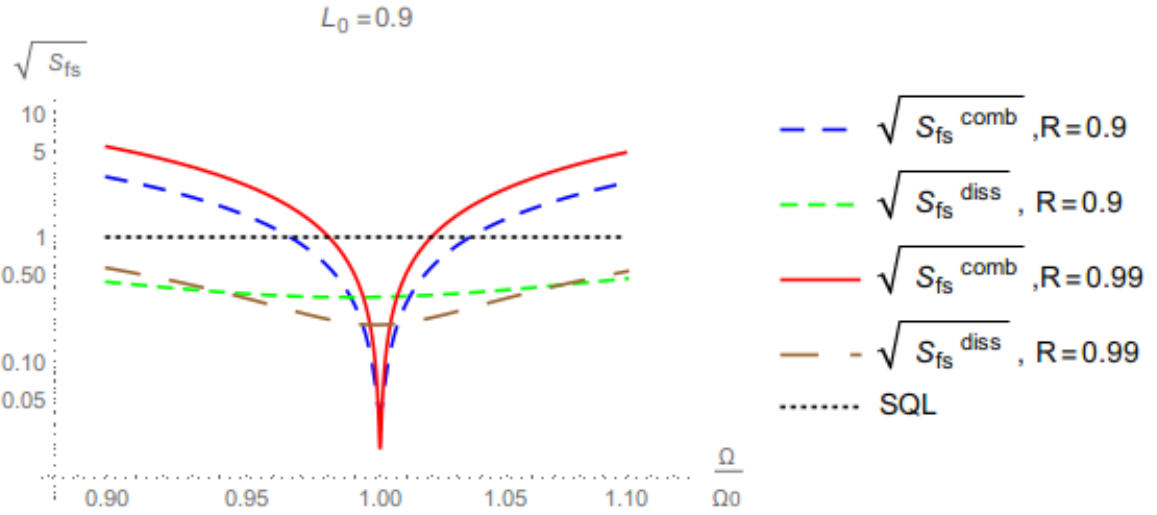}
 \caption{Graphs of amplitude spectral densities $\sqrt{S_{fs}^\text{comb}(\Omega)}$ and $\sqrt{S_{fs}^\text{diss}(\Omega)}$ with the same pump power ($\hslash k A^2$ is constant) plotted for  parameter $L_0=0.75$ (top) and $L_0=0.9$ (bottom)  for different reflectivities $R_0$. }\label{Plots1}
\end{figure*}

At first sight it looks strange, because squeezed quadratures  in case of pure dispersive and pure dissipative coupling are strongly different --- see Fig.~\ref{Pplane}. However, the key feature is the entanglement between reflected and transmitted light due to back action. Obviously, measurement of only transmitted light (or only reflected one) can not surpass SQL. Only right combination of correct chosen quadratures in reflected and transmitted light allow to cancel back action (completely on previously chosen spectral frequency) and to surpass SQL.

 Formulas for PSD $S_{fs}^\text{comb}$ and $S_{fs}^\text{sizes}$ are quite similar but not identical. The difference deals with the fact that for the same pump power one can get larger sensitivity (less  $S_{fs}^\text{comb}$) for combined coupling than for dissipative one. The physical reason  is associated with the optical rigidity \eqref{Kfs} caused by combined coupling. In the case of a purely dissipative (dispersion) coupling the sensitivity can be increased (in a lower bandwidth) by increasing the recalculated pump power $\mathcal K_0$ \eqref{Sfs}. In this case the best sensitivity indicators are achieved at $R_0\to 1$. In case of combined coupling a resonant frequency appears, defined by optical rigidity \eqref{Kfs}. By selecting the frequency $\Omega_0$ near the resonant frequency (or $(L_0-1)^2 \ll 1$) and slightly shifting it, PSD $S_{fs}^\text{comb}$ can be drastically changed at the same optical pump (i.e. $\hslash k A^2$ is constant). This can lead to a large gain in sensitivity (smaller $S_{fs}^\text{comb}$) relative to the case of dissipative (or dispersive) coupling, as shown in the Fig.~\ref{Plots1}. Indeed, comparing (\ref{KK}, \ref{Sfs})  with  (\ref{Kfs}, \ref{SfsMin}) we see that the minimum PSD $S_{fs}^\text{diss}(\Omega_0)$ is achieved for dissipative coupling and $S_{fs}^\text{comb}(\Omega_0)$ at the same power if
 \begin{align}
  \frac{2(L_0-1)^2}{R_0^2(1+R_0^2)} = 1.
 \end{align}
  For $R_0\simeq 1$ it means $L_0\simeq 2$. 
  
  In Fig.~\ref{Plots1} we presents
  graphs of amplitude spectral densities $\sqrt{S_{fs}^\text{comb}(\Omega)}$ for combined coupling and $\sqrt{S_{fs}^\text{diss}(\Omega)}$ for dissipative one with the same pump power (i.e. $\hslash k A^2$ is constant). We do not present graphs for dispersive coupling because corresponding formulas practically coincide with each others in case $R_2\to 1$; indeed, \eqref{b1ap} is similar to \eqref{a1ap} and they coincide at $R_2\to 1$. The graphs for combined coupling are  plotted for different parameters $L_0$ and $R_0$. Recall, parameter $L_0$ characterizes closeness of preliminary chosen frequency $\Omega_0$ to resonant frequency, see definition \eqref{L0}. We see that sensitivity curves for $L_0=0.9$ are more deep but more narrow in comparison with $L_0=0.75$ the same dependence is on $R_0$. It is in agreement with \eqref{DeltaOmega}: minimum of sensitivity as well as bandwidth are proportional to $\sim T_0(L_0-1)^2$.

\section{Conclusion}

We analyzed the cavity-less variants of dissipative and dispersive opto-mechanical coupling and showed that in case of dissipative coupling information on mechanical displacement as well as on the back action is provided by {\em amplitude} quadratures of reflected and transmitted waves. In contrast, in case of dispersive coupling the information on displacement is contained in {\em phase} quadrature of reflected wave only, see Fig.~\ref{Pplane}.

We considered combined coupling based on the Michelson-Sagnac interferometer (Fig.~\ref{MSI}) when both dissipative and dispersive  coupling took place. For simplicity we considered a case without cavity having only one pump. 
The main feature of combined coupling is the optical rigidity \eqref{Kfs} which appears as a result of both kinds of coupling.
 
Although back action is acting differently in case of pure dissipative and pure dispersive coupling (as illustrated in Fig.~\ref{Pplane}) we showed that variation measurement \cite{93a1VyMaJETP, 95a1VyZuPLA, 02a1KiLeMaThVyPRD} can be applied to the case of combined coupling. The reason is the entanglement between transmitted and reflected light. Obviously, measuring only transmitted (or only reflected) light we can not surpass SQL, only their correct combination allows to cancel back action  and to overcome SQL in bandwidth close to previously chosen frequency $\Omega_0$.

Moreover, at the same pump power SQL is surpassed more easily for combined coupling than in case of pure dissipative (or dispersive) coupling. The physical reason behind it deals with the optical rigidity introduced by combined coupling. 

Note that combined coupling requires more complicated measuring procedure involving homodyne detection of {\em both} reflected and transmitted waves followed by taking the optimal sum.

We would like to emphasize that we explored the cavity-less case of combined coupling. The case of combined coupling {\em with} cavity should be investigated separately since in this case we can use only one reflected wave (the end mirror is assumed to be perfectly reflecting). For example, pure dissipative coupling analyzed in this paper provides for a displacement quantum transducer, whereas pure dissipative coupling in cavity acts as a quantum speed meter  \cite{16a1PRAVyMa} rather than displacement meter.

\acknowledgments
 The authors are grateful for support provided by the Russian Foundation for Basic Research (Grant No. 19-29-11003) and for TAPIR GIFT MSU Support from the California Institute of Technology. This document has LIGO number P2000155.

\appendix

\section{Analysis of MSI}\label{appMSI}

Here we analyze MSI shown as a dashed rectangle in Fig.~\ref{MSI} with a 50/50 moving BS (having coordinate $y$), a moving completely reflecting mirror $M$ (having  coordinate $x$). We calculate input-output relations and the Lebedev forces affecting mirror $M$ and BS. 

Complex wave amplitudes $\mathcal A,\ \mathcal A_1,\ \mathcal B,\ \mathcal B_1, \mathcal A_{n,e}, \mathcal B_{n,e}$ are taken on non-shifted BS. 

\paragraph*{Input-output relations.} We start with equations:
\begin{subequations}
\label{initMSI}
 \begin{align}
 \label{Aen}
  \mathcal A_e &= \frac{i\mathcal B + \mathcal Ae^{-i\sqrt 2 ky_{bs}}}{\sqrt 2},\quad 
  \mathcal A_n = \frac{\mathcal Be^{i\sqrt 2 ky_{bs}}+i \mathcal A}{\sqrt2},\\
  \mathcal A_1 & = \frac{\mathcal B_e e^{-i\sqrt 2 ky_{bs}} + i\mathcal B_n}{\sqrt 2},\quad 
    \mathcal B_1 = \frac{ i\mathcal B_e + \mathcal B_n e^{i\sqrt 2 ky_{bs}}}{\sqrt 2}\nonumber
  \end{align}
  Here coordinates $x_m,\ y_{bs}$ denote position of mirror M and BS. We have for waves incident on BS from inside MSI 
  \begin{align}
    \label{Ben}
      \mathcal B_e &=  \mathcal A_e e^{-2ikx_m } e^{2i\phi_e},\quad 
      \mathcal B_n =  \mathcal A_n e^{2ikx_m }e^{2i\phi_n} ,
  \end{align}
  \end{subequations}
  where constants $\phi_{e,n}$ describe the phase gain in wave propagating from BS to mirror $M$ when position of BS is $y_{bs}=0$ and position of mirror M is $x_m=0$. Substituting \eqref{Ben} into  \eqref{Aen} and assuming $e^{2i\phi_e}= 1,\quad e^{2i\phi_n}= -1$, we obtain \eqref{MSIA1}.

%
 
\paragraph*{The Lebedev light pressure force $F_{m}$ applied to mirror $M$.}  In general case
\begin{subequations}
 \begin{align}
 F_{m} &= 2\hslash k\left( |\mathcal A_n|^2 - |\mathcal A_e|^2 \right)=\\
  \label{FbaM}
     & = 2\hslash k\,i \left(\mathcal A\mathcal B^*e^{-ik\sqrt 2 y_{bs}} - \mathcal A^*\mathcal Be^{ik\sqrt 2 y_{bs}}\right)
 \end{align}
\end{subequations}
Here we used relations \eqref{initMSI}.
 
\paragraph*{The Lebedev light pressure force acting on BS} along axes $\xi,\ \eta$ in Fig.~\ref{MSI} is equal to
\begin{subequations}
  \begin{align}
    F_{\xi BS}& = \hslash k\left(|\mathcal B|^2+|\mathcal B_1|^2 -2|\mathcal A_e|^2\right)   ,\\
    F_{\eta BS} & = \hslash k\left(|\mathcal A|^2+|\mathcal A_1|^2 -2|\mathcal A_n|^2\right).
   \end{align}
 \end{subequations}
We see that the light pressure force applied along axis $y$ is equal to
 \begin{subequations}
 \label{FyBS}
  \begin{align}
    F_{bs} & = \frac{F_{\xi BS} - F_{\eta BS} }{\sqrt 2}=\\
         \label{FyBS1}
	 &= \sqrt 2 \hslash k\left\{R^2\big[| \mathcal B|^2-|\mathcal A|^2\big] -\right.\\
	 \label{FyBS2}
	 &\quad \left. -  R T \left(\mathcal A \mathcal B^*e^{-i\sqrt 2 ky_{bs}} 
	  + \mathcal A^* \mathcal B e^{i\sqrt 2 ky_{bs}}\right)\right\}+\\
	 \label{FyBS3}
	 & \quad + \sqrt 2 \hslash k\, i\left\{
	   \mathcal A \mathcal B^*e^{-i\sqrt 2 ky_{bs}} - \mathcal A^* \mathcal B e^{i\sqrt 2 ky_{bs}}\right\}
  \end{align}
\end{subequations}
Here we made use of relations (\ref{initMSI}, \ref{MSIA1}) and took $y_0=0$ (defined in \eqref{xy}). 
In the light pressure force \eqref{FyBS} terms \eqref{FyBS1} and \eqref{FyBS2} correspond to {\em dispersive} coupling, whereas term \eqref{FyBS3} stands for {\em dissipative} coupling.

\section{Moving mirror}

Here we analyze  mirror $M_2$ having reflectivity $R_2$ and transmittance $T_2$ which can move as a free test mass along $z$ axis as shown in Fig.~\ref{Disp}.  We calculate input-output relations and the Lebedev forces applied to the mirror. 

Complex wave amplitudes $\mathcal A,\ \mathcal A_1,\ \mathcal B,\ \mathcal B_1$ are taken on immobile mirror. 

\paragraph*{The input-output relations} are quite obvious
\begin{subequations}
  \label{M1}
  \begin{align}
   \mathcal A_1 & =  \mathcal B T_2  + e^{2ik z} \mathcal A R_2, \\
   \mathcal B_1 &= \mathcal A  T_2- e^{-2ik z}\mathcal B R_2,
   \end{align}
 \end{subequations}
 
\paragraph*{The Lebedev light pressure force $F_{m1}$ applied to mirror $M_2$:}
\begin{subequations}
 \label{Fm1}
 \begin{align}
   \label{Fm1a}
  F_{m1} &= \hslash k\left( |\mathcal A|^2 + |\mathcal A_1|^2 -|\mathcal B|^2 - |\mathcal B_1|^2 \right)=\\
  \label{Fm1b}
     & = 2\hslash k\, \left(R_2^2|\mathcal A|^2 - R_2^2|\mathcal B|^2+\frac{}{} \right.\\
     &\qquad \left.\frac{}{} + T_2R_2\left[\mathcal A\mathcal B^*e^{2ikz} + \mathcal A^*\mathcal Be^{-2ikz}\right]\right)
 \end{align}
\end{subequations} 
 Here we substitute \eqref{M1} into \eqref{Fm1a}.
 
 \section{Deriving \eqref{SfsMin}}\label{derivation}
 
 We would like to rewrite set of equations \eqref{MSIA1cquad} in form, containing position $y$ only in two (not three) equations.
 Let us introduce notations for input and output amplitude quadratures
 \begin{subequations}
 \label{outquadAtt}
 \begin{align}
  d_{a} &= R_0 a_{a} -T_0  b_{a},\\
  e_{a} &= - T_0 a_{a} - R_0 b_{a} \, ,\\
  g_{1a} & = T_0 a_{1a} +R_0  b_{1a}, \\ 
  j_{1a} & = R_0 a_{1a} - T_0  b_{1a}
 \end{align}
 \end{subequations}
 and  present \eqref{MSIA1cquad} in form
 \begin{subequations}
  \begin{align}
   g_{1a} & =  T_0 d_{a} +R_0  e_{a}+A2ky,\\ 
   j_{1a} & = R_0 d_{a} - T_0 e_{a},\\
    a_{1p} & =  - T_0 b_p +R_0 a_p  - R_0 A\,2k y,\\
   b_{1p} &=  - T_0 a_p -R_0 b_p,\\
   A2k y &= \frac{L  \left\{-R_0d_{a} +  b_p \right\}}{(T_0R_0)(L-1)}
		+ \sqrt\frac{2L }{T_0R_0 }\cdot\frac{f_s}{L-1},\\
	L&= \frac{T_0R_0 \, 4\hslash k^2A^2}{m_{bs}\Omega^2},\quad f_s=\frac{F_s}{\sqrt{2\hslash m_{bs}\Omega^2}} 
  \end{align}
 \end{subequations}
In equation \eqref{outquadAtt} we chose $j_{1a}$ so that $j_{1a}$ does not contain information on position $y$. The other new quadratures  are chosen so set of quadratures 
 $d_a,\, e_a,\, a_p,\, b_p$  do not correlate with each other
 forming an orthogonal basis with PSD equal to:
\begin{align}
 S_{da}&=1,\quad S_{ea}=1,\quad S_{ap}=1,\quad S_{bp}=1
\end{align}

Let us estimate the weighted sum
 \begin{subequations}
  \label{G}
  \begin{align}
   G &= C\left(g_{1a} + A_a j_{1a}\right) + D\left( a_{1p} + A_p b_{1p}\right)=\\
	 \label{G1}
	  = & \left\{ C\big[ T_0 +A_a R_0 \big] +  \frac{R_0\big(DR_0-C\big]L }{T_0R_0 (L -1)}\right \}  d_{a} -\\
	 \label{G2}
     &\quad  - \left\{ D\big[ T_0 +A_p R_0\big]+\frac{\big(DR_0-C\big]L }{T_0R_0 (L -1)} \right\} b_{p} +\\
     & \quad 
		  + C\big[R_0 -A_aT_0\big]  e_{a} +D\big[ R_0 -A_pT_0\big] a_{p}+\\
		&\qquad +\big[C-DR_0\big] \sqrt\frac{2L }{T_0R_0 }\cdot\frac{f_s}{L -1}
  \end{align}
 where $A_a,\ A_p,\, C,\, D$ are the constants to be found.
 \end{subequations}
 
Let's require that there is no back action at a given frequency $\Omega_0$. This allows finding $A_a,\ A_p$ through equating terms (\ref{G1}, \ref{G2}) to zero in \eqref{G}:
 \begin{subequations}
  \label{AaAp}
  \begin{align}
		A_a & = -\frac{1}{R_0}\left(T_0  +  \frac{\big(DR_0-C\big]L_0}{C T_0 (L_0-1)} \right),\quad 
		L_0=L(\Omega_0) \nonumber 	\\
    A_p& = -\frac{1}{DR_0}\left(DT_0 +\frac{\big(DR_0-C \big]L_0}{T_0R_0 (L_0 -1)}\right).
  \end{align}
 \end{subequations}
Now we substitute $A_a,\ A_p$ into \eqref{G}
\begin{subequations}
 \label{GG}
 \begin{align}
  G &= \frac{\big[C-DR_0\big]}{(L_0-1)} \sqrt\frac{2L_0 }{T_0R_0 }\times\\
		&\left\{
		 \sqrt\frac{T_0R_0 }{2L_0 }\left[ \frac{DR_0L_0- C}{R_0(C-DR_0)}\right]    e_{a}+\right.\\
		&\left. + \sqrt\frac{T_0R_0 }{2L_0 } \left[ \frac{2DR_0L_0 -DR_0- CL_0 }{R_0^2(C-DR_0)} \right] a_{p}
		 +f_s
		\right\} 
 \end{align}
\end{subequations}
 and calculate PSD normalized to SQL
 \begin{subequations}
  \begin{align}
   S_{fs}(\Omega_0) &= \frac{T_0}{2R_0 (C -DR_0)^2}\times\\
       &\left\{\left[DR_0\sqrt L_0 - \frac{C}{\sqrt L_0}\right]^2 + \right.\\
		&\qquad \left. + \left[\frac{ (2DR_0 -C )\sqrt L_0}{R_0}  - \frac{D}{\sqrt L_0}\right]^2\right\}
   \end{align}
  \end{subequations}
 Let's choose $D,\ C$  under which $S_{fs}(\Omega_0)$ attains the minimum. Obviously $S_{fs}(\Omega_0)$ depends only on the ratio $D/C$: 
 \begin{align}
	\left.\frac{D}{C}\right|_{opt}& = \frac{L_0 + R_0^2}{R_0 \left(L_0 \big[R_0^2+2 \big]-1\right)}.	 
 \end{align}

Thus, for arbitrary frequency $\Omega$ PSD $S_{fs}^{min}$ equals to \eqref{SfsMin} if we assume $D/C=(D/C)_{opt}$.


\end{document}